\documentstyle[12pt]{article}
\newlength{\extraspace}
\newlength{\extraspaces}
\newcommand{\be}{\begin{equation}
\addtolength{\abovedisplayskip}{\extraspaces}
\addtolength{\belowdisplayskip}{\extraspaces}
\addtolength{\abovedisplayshortskip}{\extraspace}
\addtolength{\belowdisplayshortskip}{\extraspace}}
\newcommand{\ee}{\end{equation}}

\newcommand{\ba}{\begin{eqnarray}
\addtolength{\abovedisplayskip}{\extraspaces}
\addtolength{\belowdisplayskip}{\extraspaces}
\addtolength{\abovedisplayshortskip}{\extraspace}
\addtolength{\belowdisplayshortskip}{\extraspace}}
\newcommand{\ea}{\end{eqnarray}}

\newcommand{\nonu}{\nonumber \\[.5mm]}
\newcommand{\A}{&\!\!\!}

\newcommand{\newsection}[1]{
\vspace{7mm} \pagebreak[3] \addtocounter{section}{1}
\setcounter{subsection}{0} \setcounter{footnote}{0}
\begin{center}
{\large {\bf \thesection. #1}}
\end{center}
\nopagebreak
\medskip
\nopagebreak \hspace{3mm}}

\begin{document}
\begin{center}
{{\bf  Gravitational radiation fields in  teleparallel
equivalent of general relativity and their energies }}\footnote{\hspace{-0.4cm}PACS numbers: 04.20.Cv, 04.20.Fy, 04.50.-h\\
\hspace*{0.15cm}Keywords:  teleparallel equivalent of general relativity, energy-momentum tensor, Bondi mass, gravitational\hspace*{0.15cm}\\ radiation}
\end{center}
\centerline{ Gamal G.L.
Nashed\footnote{\hspace{-0.4cm}Mathematics Department, Faculty of Science, Ain
Shams University, Cairo, Egypt}}

\bigskip

\centerline{\it Centre for Theoretical Physics, The British
University in Egypt,
 El-Sherouk City,}
\centerline{{\it Misr - Ismalia Desert Road, Postal No. 11837,
P.O. Box 43, Egypt.} }

\bigskip
 \centerline{ e-mail: nashed@bue.edu.eg}

\hspace{2cm}

We derive two new  retarded solutions in the teleparallel
theory equivalent to general relativity (TEGR). One of these solutions gives a divergent  energy. Therefore, we used the regularized expression of
the gravitational energy-momentum tensor, which is a coordinate
dependent. A detailed analysis of the loss of
the mass of Bondi space-time is carried out  using the flux of the gravitational
energy-momentum.
\\
\\
\\

\begin{center}
\newsection{\bf Introduction}
\end{center}

The linearized theory has been developed extensively, however, it
seems  doubtful if its results  can be fully trusted
 \cite{SN}. The non-linearity of the gravitational field is the
most characteristic properties and some of the crucial properties
of the field  show themselves only through the non-linear terms.
Additionally, it is never clear whether solutions derived from the
linearization corresponds to exact solutions \cite{SN}. In spite
that we have a good deal about exact plane and cylindrical wave
solutions, it is doughtily whether such solutions display the
  characteristics of physically significant waves \cite{BBM}. Einstein's
General relativity (GR) theory is a complete one however, it may not
give sensible solutions for situations too far removed from what is physically reasonable.
 The most simplest field due to a finite source which can be obtained from GR is the
spherically symmetric solution, however, Birkhoff's theorem shows
that such solution must be static. For this reason, one cannot be
truly speak about spherically symmetric waves and
 thus any description of radiation from a finite system must
 necessary involve three coordinate significantly. This makes
 the mathematical calculations more complicated and thus we have to
 make use of methods of approximation.

  Two well established formulas of gravitational energy are the ADM
 (Arnowitt-Deser-Misner) energy that  corresponds to the total
 energy \cite{ADM} and the Bondi energy \cite{BBM} which describes the
 mass of radiating systems in asymptotic flat space-times. The
 loss of energy of the total mass of a source that radiates
 gravitational waves is main feature of the Bondi energy \cite{BBM}, i.e.,
\be \frac{dE}{dt}=-\left(\frac{dc_0}{dt}\right)^2,\ee  where $c_0$
is the news function. Eq. (1) shows that Bondi energy is related
to the loss of the total mass of a source that radiates
gravitational waves. The loss of the total mass is given in terms
of the minus news function squared which means that the mass of
the source can decrease.

Teleparallel theories are interesting for several reasons: first
of all, general relativity (GR) can be viewed as a particular
theory of teleparallelism and, thus, teleparallelism could be
considered at the very least as a different point of view that can
lead to the same results \cite{TM}.  Second, in this framework,
one can define an energy-momentum tensor for the gravitational
field that is a true tensor under all general coordinate
transformations. This is the reason why teleparallelism was
reconsidered by M\o ller  when he was studying the problem of
defining an energy-momentum tensor for the gravitational field
\cite{M2}. The idea was taken over by Pellegrini and Pleba\'{n}ski
that constructed the general Lagrangian for these theories
\cite{PP}. The third reason why these theories are interesting is
that they can be seen as gauge theories of the translation group
(not the full Poincar´e group) and, thus, they give an alternative
interpretation of GR \cite{Hw,NHC,AGP,HS}. Therefore, we consider the tetrad theory of
 gravitation in this work.

It is the aim of the present work to study the variation of the
gravitational energy and the corresponding total flux of energy at
spacelike infinity within the framework of TEGR. To do so we
derive  the solution given by Bondi \cite{BBM} in the
framework of TEGR and then compute the total mass using the
energy-momentum tensor. Also we compute the loss of mass using the
flux of the gravitational energy-momentum tensor \cite{MDTC}. In \S 2, we
briefly review the TEGR theory for gravitational field and  its derivation of
the field equations. A summary of the derivation of energy and
total fluxes of the gravitational energy-momentum  are also given in \S 2. In \S 3,
we study the most general tetrad field which contains 16 unknown
functions. Two {\it new} solutions are also given in \S 3. Then, we
calculate the total mass of these solutions using the regularized
expression in \S 4. Using the flux of the gravitational
energy-momentum we also calculate the loss of mass in \S 4. Final
section is devoted for the main results.

\newsection{The continuity equation and the fluxes of the
 gravitational and matter fields}
In a space-time with absolute parallelism the parallel vector
fields ${e_a}^\mu$ define the nonsymmetric affine connection \be
{\Gamma^\lambda}_{\mu \nu} \stackrel{\rm def.}{=} {e_a}^\lambda
{e^a}_{\mu, \nu}, \ee where $e_{a \mu, \ \nu}=\partial_\nu e_{a
\mu}$\footnote{space-time indices $\mu, \ \ \nu, \cdots$ and
SO(3,1) indices a, b $\cdots$ run from 0 to 3. Time and space
indices are indicated to $\mu=0, i$, and $a=(0), (i)$.}. The
curvature tensor defined by ${\Gamma^\lambda}_{\mu \nu}$, given by
Eq. (2), is identically vanishing. The metric tensor $g_{\mu \nu}$
 is defined by
 \be g_{\mu \nu} \stackrel{\rm def.}{=}  \eta_{a b} \ {e^a}_\mu {e^b}_\nu, \ee
with $\eta_{a b}=(-1,+1,+1,+1)$ is the metric of Minkowski
space-time.

  The Lagrangian density for the gravitational field in the TEGR,
  in the presence of matter fields, is given by\footnote{Throughout this paper we use the
relativistic units$\;$ , $c=G=1$ and $\kappa=8\pi$.} \cite{MDTC} \be
{\cal L}_G  =  e L_G =- \displaystyle {e \over 16\pi}  \left(
\displaystyle {T^{abc}T_{abc} \over 4}+\displaystyle
{T^{abc}T_{bac} \over 2}-T^aT_a
  \right)-L_m= - \displaystyle {e \over 16\pi} {\Sigma}^{abc}T_{abc}-L_m,\ee
where $e=det({e^a}_\mu)$. The tensor ${\Sigma}^{abc}$ is defined
by \be {\Sigma}^{abc} \stackrel {\rm def.}{=} \displaystyle{1
\over 4}\left(T^{abc}+T^{bac}-T^{cab}\right)+\displaystyle{1 \over
2}\left(\eta^{ac}T^b-\eta^{ab}T^c\right).\ee $T^{abc}$ and $T^a$
are the torsion tensor and the basic vector field  defined by \be
{T^a}_{\mu \nu} \stackrel {\rm def.}{=}
{e^a}_\lambda{T^\lambda}_{\mu
\nu}=\partial_\mu{e^a}_\nu-\partial_\nu{e^a}_\mu, \qquad \qquad
{T^a}_{b c} \stackrel {\rm def.}{=}  {e_b}^\mu {e_c}^\nu
{T^a}_{\mu \nu} \qquad \qquad T^a \stackrel {\rm
def.}{=}{{T^b}_b}^a.\ee The quadratic combination
$\Sigma^{abc}T_{abc}$ is proportional to the scalar curvature
$R(e)$, except for a total divergence term \cite{MDTC}. $L_m$
represents the Lagrangian density for matter fields.

The gravitational  field equations described by ${\it L_G}$ are
the following
 \be  e_{a \lambda}e_{b \mu}\partial_\nu\left(e{\Sigma}^{b \lambda \nu}\right)-e\left(
 {{\Sigma}^{b \nu}}_a T_{b \nu \mu}-\displaystyle{1 \over 4}e_{a \mu}
 T_{bcd}{\Sigma}^{bcd}\right)= \displaystyle{1 \over 2}{\kappa} eT_{a
 \mu},\ee
where \[ \displaystyle{ \delta L_m \over \delta e^{a \mu}} \equiv
e T_{a \mu}.\] It  is possible to prove by explicit calculations
that the left hand side of the symmetric field equations (7) is
exactly given by \cite{MDTC}
 \[\displaystyle{e \over 2} \left[R_{a
\mu}(e)-\displaystyle{1 \over 2}e_{a \mu}R(e) \right]. \]
Multiplication of  Eq. (7) by the appropriate inverse tetrad
fields yields it to have the form \cite{MDTC,MVR}  \be
\partial_\nu \left(-e {\Sigma}^{a \lambda \nu } \right)=-\displaystyle{e
e^{a \mu} \over 4} \left(4{\Sigma}^{b \lambda \nu }T_{b \nu \mu }-
{\delta^\lambda}_\mu {\Sigma}^{bdc}T_{bcd} \right)-4\pi {e^a}_\mu
T^{\lambda \mu}.\ee By restricting the space-time index $\lambda$
to assume only spatial values then Eq. (8) takes the form
\cite{MDTC} \be \partial_0 \left(e {\Sigma}^{a 0 j}
\right)+\partial_k\left(e {\Sigma}^{a k j}
\right)=-\displaystyle{e e^{a \mu} \over 4}\left(4{\Sigma}^{b c
j}T_{b c \mu }- {\delta^j}_\mu {\Sigma}^{bcd}T_{bcd} \right)-4\pi
e {e^a}_\mu T^{j \mu}.\ee Note that the last two indices of
${\Sigma}^{abc}$ and $T^{abc}$ are anti-symmetric. Taking the
divergence of Eq. (9) with respect to j yields

\be -\partial_0 \partial_j \left(-\displaystyle {1  \over 4 \pi} e
{\Sigma}^{a 0 j} \right)=-\displaystyle{1 \over 16 \pi}
\partial_j\left[e e^{a \mu}\left(4{\Sigma}^{b c j}T_{b c \mu }-
{\delta^j}_\mu {\Sigma}^{bcd}T_{bcd} \right)-16 \pi( e {e^a}_\mu
T^{j \mu})\right].\ee

The Hamiltonian formulation of TEGR is obtained by establishing
the phase space variables. The Lagrangian density does not contain
the time derivative of the tetrad component $e_{a0}$. Therefore,
this quantity will arise as a Lagrange multiplier \cite{Dp}. The
momentum canonically conjugated to $e_{ai}$ is given by
$\Pi^{ai}=\delta L/\delta \dot{e}_{ai}$. The Hamiltonian
formulation is obtained by rewriting the Lagrangian density in the
form $L=p \ \dot{q}-H$, in terms of $e_{ai}, \Pi^{ai}$ and the
Lagrange  multipliers. The Legendre transformation can be
successfully carried out and the final form of the Hamiltonian
density has the form \cite{MR} \be H=e_{a0}C^a+\alpha_{ik}
\Gamma^{ik}+\beta_k\Gamma^k,\ee plus a surface term. Here
$\alpha_{ik}$ and $\beta_k$ are Lagrange multipliers that are
identified as \be \alpha_{ik}={1 \over 2} (T_{i0k}+T_{k0i}) \qquad
and \qquad \beta_k=T_{00k},\ee where $C^a$, $\Gamma^{ik}$ and
$\Gamma^k$ are first class constraints.

 The constraint $C^a$ is
written as $C^a=-\partial_i \Pi^{ai}+h^a$, where $h^a$ is an
intricate expression of the field variables. The integral form of
the constraint equation $C^a=0$ motivates the definition of the
gravitational energy-momentum $P^a$ four-vector \cite{MDTC} \be
P^a=-\int_V d^3 x
\partial_i \Pi^{ai},\ee where $V$ is an arbitrary volume of the
three-dimensional space. In the configuration space we have \ba
\Pi^{ai} \A =\A -4\kappa \sqrt{-g} \Sigma^{a0i} \quad with \quad
\partial_\nu(\sqrt{-g}\Sigma^{a \lambda \nu})=\displaystyle{1
\over 4 \kappa}\sqrt{-g}{e^a}_\mu (t^{\lambda \mu}+T^{\lambda
\mu}) \quad where \nonu
\A \A   t^{\lambda \mu}=\kappa \left(4\Sigma^{bc
\lambda}{T_{bc}}^\mu-g^{\lambda \mu} \Sigma^{bcd}T_{bcd}
\right).\ea

By integrating Eq. (10) on a volume $V$ of the three-dimensional
space  we get \cite{MDTC} \be \frac{d}{dt}\left[-\int_v d^3x
\partial_j \Pi^{aj} \right]=-{\phi_g}^a-{\phi_m}^a,\ee where
\be {\phi_g}^a=\frac{\kappa}{2}\int_S dS_j \left[ e e^{a
\mu}\left(4\Sigma^{bcj}T_{bc\mu}-{\delta_\mu}^j
\Sigma^{bcd}T_{bcd}\right)\right],\ee is the {\it a} component of
the gravitational energy-momentum flux and \be {\phi_m}^a=\int_S
dS_j\left(e {e^a}_\mu T^{j \mu}\right), \ee is the {\it a}
component of the matter energy-momentum flux. $S$ represent the
spatial boundary of the volume $V$. Therefore the loss of the
gravitational energy $P^{(0)}=E$ is governed by the equation \be
\frac{dE}{dt}=-{\phi_g}^0-{\phi_m}^0.\ee Maluf and Faria have
shown that ${\phi_g}^0$ played a major role to the Bondi's
radiating space-time, since it yields the well known expression
for the gravitational energy loss, while ${\phi_m}^0$ is related
to the Vaidya's space-time \cite{MF}.

\newsection{Bondi space-times in TEGR and their associated energies}
In this section we are going to use the most general tetrad field
which contains 16 unknown functions. Let us write it in the
coordinate $(t, \ r,\ \theta,\ \phi)$  by \be \left({e^a}_\mu
\right) =\left(\matrix {A_1& A_2 & A_3 & A_4\vspace{3mm} \cr B_1&
B_2 & B_3 & B_4 \vspace{3mm} \cr F_1&F_2&F_3&F_4 \vspace{3mm} \cr
H_1&H_2&H_3&H_4 \cr } \right), \ee where $A_i, \ B_i,  \ F_i, \
H_i $, $i=1\cdots 4$ are unknown functions of $r$, $\theta$,
$\phi$ and $t$. Applying Eq. (19) to the field equations (7) we
obtained a very lengthy and tedious  system of partial
differential equations. We are going to write a special
solutions to these system.\\ \underline{\it First solution:}\\
The 16 unknown functions   in this solution have the form \ba A_1
\A=\A -e^\beta \sqrt{\frac{V}{r}}, \qquad A_2 = -e^\beta
\sqrt{\frac{r}{V}}, \qquad A_3=0, \qquad  A_4=0, \qquad B_1 =
 U r e^\gamma \cos{\theta}\cos{\phi}, \nonu
 B_2 \A=\A -e^\beta \sqrt{\frac{r}{V}} \sin{\theta}\cos{\phi}, \qquad
B_3=-re^\gamma \cos{\theta}\cos{\phi}, \qquad B_4=-re^{-\gamma}
\sin{\theta}\sin{\phi},\nonu
F_1 \A=\A U r e^\gamma  \cos{\theta}\sin{\phi}, \qquad F_2 =
-e^\beta \sqrt{\frac{r}{V}} \sin{\theta}\sin{\phi}, \qquad
F_3=-re^\gamma \cos{\theta}\sin{\phi}, \nonu
 F_4 \A=\A re^{-\gamma} \sin{\theta}\cos{\phi},  \qquad \qquad H_1 =
-U r e^\gamma \sin{\theta} ,  \qquad \qquad H_2 = -e^\beta
\sqrt{\frac{r}{V}} \cos{\theta}, \nonu
H_3\A=\A re^\gamma \sin{\theta},  \qquad \qquad  \quad \qquad
H_4=0. \ea The functions $\beta$, $\gamma$, $V$ and  $U$ appear in
Eq. (20) have the following asymptotic behavior up to
$O\left(\frac{1}{r^2}\right)$ \ba \beta \A=\A
-\frac{\left(c^2(u,\theta)+d^2(u,\theta)\right)}{4r^2}+O\left(\frac{1}{r^4}\right),
\qquad \gamma=\frac{c(u,\theta)}{r}+O\left(\frac{1}{r^3}\right),
\nonu
U\A=\A -\frac{l(u,\theta)}{r^2}+O\left(\frac{1}{r^3}\right),
\qquad V=r-2M(u,\theta)-\frac{l_1(u,\theta)}{r},\ea where the time
$t$ is defined as $t=u+r$ which is the retarded time. The
quantities $l(u,\theta)$ and $l_1(u,\theta)$ are defined as \ba
l(u,\theta) \A=\A\frac{\partial c(u,\theta)}{\partial
\theta}+2c(u,\theta) \cot{\theta}, \nonu
l_1(u,\theta)\A=\A \frac{\partial d(u,\theta)}{\partial
\theta}+d(u,\theta) \cos{\theta}-\left(\frac{\partial
c(u,\theta)}{\partial
\theta}\right)^2-4c(u,\theta)\left(\frac{\partial
c(u,\theta)}{\partial \theta}\right)\cot{\theta}\nonu
\A \A -\frac{1}{2}c^2(u,\theta)\left(1+8\cot^2{\theta}\right),\ea
with $M(u,\theta)$, $c(u,\theta)$ and $d(u,\theta)$ are the mass
aspect, the news function and the dipole aspect respectively.
Solution given by Eq. (20) with Eqs. (21) and (22) satisfies the
field equations (7) up to $O\left(\frac{1}{r^2}\right)$. The
metric associated with solution given by Eq. (20) has the form \ba
ds^2\A=\A-\left(\frac{V}{r}e^{2\beta}-U^2r^2e^{2\gamma}\right)du^2-2e^{2\beta}du
dr\nonu
\A \A -2Ur^2e^{2\gamma}du
d\theta+r^2\left(e^{2\gamma}d\theta^2+e^{2\gamma}\sin^2{\theta}d\phi^2\right),\ea
which is the Bondi space-time \cite{MDTC}.\\
\underline{\it Second solution:}\\
The 16 unknown functions in this case have the form \ba A_1 \A=\A
\sqrt{{\frac {V  {e}^{2\,\beta }}{r}}+{r}^{2}{e}^{2\,\gamma} U^{2}
\cosh  2\,\delta
 +{r}^{2 }{e}^{-2\,{\gamma} } W^{2}\cosh  2\,\delta +2\,{r}^{2}W  U
  \sinh 2\,\delta}, \nonu
A_2\A=\A-B_2 = {\frac {-{e}^{2\,\beta }}{\sqrt {{ \frac {V
{e}^{2\,\beta }}{r}}+{r}^{2}{e}^{2\,{ \gamma}  }  U^2 \cosh
2\,\delta   +{r}^{2 }{e}^{-2\,{\gamma}  }  W^2 \cosh 2\,\delta
+2\,{r}^{2}W U  \sinh 2\,\delta
  }}}, \nonu
 A_3\A=\A-B_3=  {
\frac {-{r}^{2} \left( {e}^{2\,{\gamma}  }U
  \cosh  2\,\delta +W \sinh
 2\,\delta  \right)}{\sqrt {{\frac {V  {e}^{2\, \beta
}}{r}}+{r}^{2}{e}^{2\,{\gamma}  }U^{2}\cosh  2\,\delta
+{r}^{2}{e}^{-2\,{\gamma}}  W^{2}\cosh
 2\,\delta +2\,{r}^{2}W
 U \sinh
 2\,\delta}}}, \nonu
  A_4\A=\A-B_4 = {\frac {-{r}^{2}\sin  \theta
\left( {e}^{-2\,{\gamma}
  }W  \cosh
 2\,\delta +U \sinh 2\,\delta   \right)}{\sqrt {{\frac {V
  {e}^{2\,\beta
}}{r}}+{r }^{2}{e}^{2\,{\gamma}  }  U^{2}\cosh  2\,\delta
+{r}^{2}{e}^{-2\,{\gamma}
  } W^{2}\cosh 2\,\delta +2\,{r}^{2}W  U
  \sinh 2\,\delta }}},  \nonu
 B_1 \A =\A F_1=F_2=H_1 = H_2 =H_3=0,\quad F_3 = r{e}^{\gamma}\sqrt{-\cosh
 2\,\delta},   \nonu
 F_4 \A =\A-{\frac {r \sinh 2\,\delta
  \sin \theta }{{e}^{\,{\gamma}
 }\sqrt {-\cosh 2\,\delta }}}, \quad H_4 = \sqrt {-{r}^{2}{e}^{-2\,{\gamma}  }
\cosh 2\,\delta
 \sin^{2} \theta +{\frac {{r}^{2}
 \sinh^{2}  2\,\delta  \sin ^{2} \theta}
{{e}^{2\,{\gamma} }\cosh 2\, \delta }}}.  \ea The functions
$\beta$, $\gamma$, $V$,\, $U$,\,  $W$ and $\delta$ appear in Eq.
(24) have the following asymptotic behavior up to
$O\left(\frac{1}{r^2}\right)$ \ba \beta \A=\A
-\frac{\left(c^2(u,\theta)+d^2(u,\theta)\right)}{4r^2}+O\left(\frac{1}{r^4}\right),
\qquad \gamma=\frac{c(u,\theta)}{r}+O\left(\frac{1}{r^3}\right),
\quad \delta={\frac {{\it d} \left( \theta, u\right)
}{r}}+O\left(\frac{1}{r^3}\right) \nonu
U\A=\A -\frac{l(u,\theta)}{r^2}+O\left(\frac{1}{r^3}\right),
\qquad V=r-2M(u,\theta)-\frac{l_1(u,\theta)}{r}, \quad W =
-\frac{l_2(u,\theta)}{r^2}+O\left(\frac{1}{r^3}\right).\nonu
\A \A \ea The quantities $l(u,\theta)$ and $l_1(u,\theta)$ are
defined in Eq. (22) and $ l_2(u,\theta)$ has the form \ba
l_2(u,\theta) \A=\A\frac{\partial d(u,\theta)}{\partial
\theta}+2d(u,\theta) \cot{\theta}.\ea The metric associated with
solution given by Eq. (24) has the form \ba
ds^2\A=\A\left(\frac{V}{r}e^{2\beta}+r^2U^2e^{2\gamma} \cosh 2\,
\delta+r^2W^2e^{-2\gamma} \cosh 2\, \delta +2r^2UW \sinh 2\,
\delta \right)du^2-2e^{2\beta}du dr\nonu
\A \A -2r^2\left(Ue^{2\gamma}\cosh 2\, \delta+W\sinh 2\,
\delta\right) du d\theta-2r^2\sin \theta \left(We^{-2\gamma}\cosh
2\, \delta+U\sinh 2\, \delta\right) du d\phi\nonu
\A \A+r^2\Biggl(e^{2\gamma}\cosh 2\, \delta\,  d\theta^2
+e^{-2\gamma}\cosh 2\, \delta \sin^2{\theta}\, d\phi^2+2\sinh 2\,
\delta \sin\theta\, d\theta d\phi \Biggr),\ea which is the Bondi
space-time \cite{Zh}. Solutions (20) and (24) satisfy the field
equations (7) up to $O\left(\frac{1}{r^3}\right)$.

Calculate the energy associated with solution (20) by
calculating the torsion tensor and basic vector\footnote{ The details components
of those quantities are given in the appendix.},  we get finally  up to $O\left(\displaystyle\frac{1}{r^0}\right)$
\be E \cong \frac{1}{2}{\int_0}^\pi \sin \theta
\,\left\{M(u,\theta)-2r\right\}\, d\theta,\ee which depends on the
radial coordinate. Therefore, in this case we are going to use the
regularized expression for the gravitational energy-momentum
\cite{MDTC}.
\newsection{Regularized expression for the gravitational energy-momentum
  and localization of energy}

For a space-time in which the physical quantities, i.e.,
$M(u,\theta)$ etc.  equal to zero, it is usual to consider a set
of tetrad fields such that the torsion ${T^\lambda}_{\mu \nu}=0$
{\it in any coordinate system}. However, in general an arbitrary
set of tetrad fields  do not satisfy ${T^\lambda}_{\mu \nu}=0$. It
might be argued, therefore, that the expression for the
gravitational energy-momentum (13) is restricted to particular
class of tetrad fields, namely, to the class of frames such that
${T^\lambda}_{\mu \nu}=0$. If ${E_a}^\mu$ represents the
space-time of the tetrad field in which  the physical quantities
equal zero \cite{MVR}. To explain this, let us calculate the
space-time in which  the physical quantities equal zero  for the
tetrad field of Eq. (19) using (20) which is given by \be
\left({{E}_a}^\mu \right) =\left(\matrix {-1&-1 &0 &0 \vspace{3mm}
\cr 0&-\sin\theta \cos \phi &r\cos\theta \cos \phi& -r\sin\theta
\sin \phi \vspace{3mm} \cr 0& -\sin\theta \sin \phi&r\cos\theta
\sin\phi&r\sin\theta \cos \phi \vspace{3mm} \cr 0&-\cos\theta
&-r\sin\theta&0 \cr } \right). \ee Expression (29) yields the
following non-vanishing torsion components: \be
{T^{2}}_{12}=\frac{2}{r}={T^{3}}_{13}.\ee

 We will denote ${T^a}_{\mu \nu}(E)=\partial_\mu {E^a}_\nu-\partial_\nu
 {E^a}_\mu$ and $\Pi^{a j}(E)$ as the expression of $\Pi^{a j}$
 constructed out of the  tetrad ${E^a}_\mu$. {\it The
 regularized form of the gravitational energy-momentum $P^a$ is
 defined by} \cite{MVR,MR}
 \be P^a=-\int_{V} d^3x \partial_k \left[ \Pi^{a k}(e)-\Pi^{a k}(E)
 \right].\ee  The reference space-time is
 determined by tetrad fields ${E^a}_\mu$, obtained from
 ${e^a}_\mu$ by requiring the vanishing of the physical parameters
 like mass, angular-momentum, etc.  Eq. (31) can have the form \cite{MR,MVR}
\be P^a=-\oint_{S\rightarrow \infty} dS_k \left[ \Pi^{a
k}(e)-\Pi^{a k}(E) \right],\ee where the surface $S$ is
established at spacelike infinity. Eq. (32) transforms as a vector
under the global SO(3,1) group \cite{MDTC}.

Let us use Eq. (32) to calculate the gravitational energy of Eq.
(20). We need to calculate the quantity  \[ \Sigma^{(0)
01}={e^{(0)}}_0\Sigma^{0 01}+{e^{(0)}}_a\Sigma^{a 01}.\]
 Evaluate  the above equation
we find \be \Pi^{(0)1}(e)\cong \frac{1}{4}\sin \theta
\left\{M(u,\theta)-4r\right\}.\ee  The expression of
$\Pi^{(0)1}(E)$ is obtained by just making $M(u,\theta)=0$, in Eq.
(33), it is given by \be \Pi^{(0)1}(E)\cong -r\sin \theta.\ee
Using Eqs. (33) and (34) in Eq. (32) we finally get the
gravitational energy of the tetrad field of Eq. (19) using
solution given by Eq. (20) in the form \be P^{(0)}\cong
\frac{1}{2} {\int_0}^\pi \sin \theta\, M(u,\theta)\, d\theta,\ee
which is the Bondi energy \cite{BBM}.

The flux of the gravitational energy which is determined by
putting $a=(0)$ in Eq. (16) and by making $j=1$, i.e., by
integrating over a surface $S$ of constant radius $r$, and
requiring $r \, \rightarrow \, \infty$. Eq. (16) in such case has
the form \be {\phi_g}^{(0)} = \frac{\kappa}{2}\int_S dS_1 e
\left(4e^{(0) \mu}\Sigma^{bc1}T_{bc\mu}-e^{(0) 1}
\Sigma^{bcd}T_{bcd}\right). \ee Applying Eq (36) to the tetrad
field (19) using Eq. (20),  we finally get \be
{\phi_g}^{(0)}=\frac{1}{2}{\int_0}^\pi \sin \theta\,
\left\{c_{,\,0}(u,\theta)\right\}^2\, d\theta,\ee which is the
value of the loss of the gravitational mass of the Bondi
space-time \cite{BBM}. In that case Eq. (1) is satisfied for
solution given by Eq. (20).

 Using the regularized form of the energy given by Eq. (32) to calculate the gravitational energy of  Eq (24), we finally
get  \be \Pi^{(0)1}(e)\cong \frac{1}{4}\sin \theta
\left\{M(u,\theta)-r\right\}.\ee The expression of $\Pi^{(0)1}(E)$
is obtained by just making $M(u,\theta)=0$, in Eq. (38), it is
given by \be \Pi^{(0)1}(E)\cong -r\sin \theta.\ee Thus the
gravitational energy of  Eq. (24) is given by \be P^{(0)}\cong
\frac{1}{2} {\int_0}^\pi \sin \theta\, M(u,\theta)\, d\theta,\ee
which is the Bondi energy \cite{BBM}.

Repeat the same calculations done for the first solution we
finally, get the flux of the gravitational energy of the second
solution in the form \be {\phi_g}^{(0)}=\frac{1}{2}{\int_0}^\pi
\sin \theta\,
\left(\left\{c_{,\,0}(u,\theta)\right\}^2+\left\{d_{,\,0}(u,\theta)\right\}^2\right)\,
d\theta,\ee which is the value of the loss of mass of the Bondi
space-time \cite{BBM,Zh}. The flux given by Eq. (41) will coincide
with that of Eq. (37) when the function $d(u,\theta)=0$ which
means that the function $\delta$ appears in solution (24) must
vanish.
\newsection{Main results and Discussion}

The main results of the present paper are the
following:\vspace{0.3cm}\\
 $\bullet$ We have obtained two new special
solutions which satisfy the field equations (7). These
solutions reproduce the Bondi space-time \cite{BBM,MDTC,Zh}.\vspace{0.3cm}\\
$\bullet$ We have calculated the energy associated with the first solution and
 found that it is depends on the radial
coordinate. This is due to the fact if the physical quantities, i.e., $M(u,\theta)$
 set equal zero we show that the components of
the torsion tensor did not vanishing as it should be! Therefore,
we have used the regularized expression  to
recalculate the energy associated with the two solutions. We
have shown that  the formula of energy associated with these
solutions
are coincide with that given by Bondi \cite{BBM}.\vspace{0.3cm}\\
$\bullet$ We have used  the flux of the energy-momentum of
gravitational and matter fields \cite{MF}. The flux of matter
field has no effect since the solutions
 are vacuum ones to the field equations. The
non-vanishing  value of the flux of the gravitational
energy-momentum  indicates the
emission of the gravitational waves.\vspace{0.3cm}\\
$\bullet$ Maluf and Faria \cite{MF} have applied the flux of the
gravitational energy-momentum to a tetrad which is
different from that we have obtained here. Our results are in coincident
with what obtained before \cite{MF}.\vspace{0.3cm}\\
$\bullet$ Zhang \cite{Zh} have discussed within the Reimannian
geometry the properties of the metric (27). Here we have discussed the
gravitational radiation of this metric within the teleparallel
geometry by using the flux of the gravitational energy-momentum.
Also we have shown that our results are in coincident  with that obtained before
 \cite{Zh}.\vspace{0.3cm}\\
$\bullet$ A further work that can be done for the  solutions
obtained  in this work is to study their spatial momentum and
angular-momentum. This will be our future work.

\bigskip
\vspace{2cm}
\centerline{\Large{\bf Appendix}}

\bigskip
\vspace{1cm}
Here in this appendix we write the explicitly the calculations used in Eq. (28) and in Eq.
(38). For Eq. (28), the non-vanishing components of torsion are \ba
{T^0}_{01}=-\frac{{T^1}_{01}}{2} \A\cong \A -2\,{\frac {M_{,\, 0}
(u, \theta ) } {r}}+O\left(\frac{1}{r^2}\right), \, \,
{T^0}_{02}\cong \frac{ {\it c_{,\, \theta}} (u, \theta )   \sin
\theta +2\,{\it c} (u,\theta ) \cos \theta - M_{,\, \theta}(u,
\theta) \sin \theta}{r \sin \theta}+O\left(\frac{1}{r^2}\right),
\nonu
 {T^0}_{12}=\frac{-{T^1}_{12}}{2} \A\cong \A 2 \,{\frac {M_{, \,\theta} (u, \theta)
 }{r}}+O\left(\frac{1}{r^2}\right), \quad  {T^1}_{02}= {T^3}_{23} \cong
  -\frac{{\it c_{,\, \theta}} ( u,\theta) \sin  \theta  +2\,{\it c}(u, \theta)
  \cos \theta}{r \sin \theta}+O\left(\frac{1}{r^2}\right),\nonu
 {T^2}_{12}={T^3}_{13}\A\cong\A\frac{2}{r}+O\left(\frac{1}{r^2}\right),
 \quad {T^3}_{03}=-\frac{c_{,\, 0}(u, \theta)}{r}+O\left(\frac{1}{r^2}\right),
 \ea
and the non-vanishing components of the basic vector are \be T^0
\cong -2\,\frac {2+M_{,\, 0}( u,\theta)
}{r}+O\left(\frac{1}{r^2}\right), \quad T^1=\frac {4+M_{,\, 0} (u,
\theta ) }{r }+O\left(\frac{1}{r^2}\right), \ee where
\[M_{,\, i}=\frac{\partial M}{\partial x^i}.\]

For Eq. (38) the non-vanishing components of the torsion tensor are
\ba {T^2}_{20}\A=\A -{T^3}_{30} \cong  -\frac{ {\it
c_{,\, 0}} (u, \theta )}{r}+O\left(\frac{1}{r^2}\right),\quad
{T^2}_{23}\cong 2\,\frac {\left\{ d(u, \theta )\cos \theta+
d_{,\,\theta}(u, \theta )\sin\theta \right\}}
{r}+O\left(\frac{1}{r^2}\right), \, \nonu
 {T^2}_{30} \A\cong \A -\,{\frac {2  d_{, \,0} (u, \theta)\, \sin \theta
 }{r}}+O\left(\frac{1}{r^2}\right), \quad  {T^3}_{23}  \cong
 \frac{r \cos  \theta  -{\it c_{,\, \theta}}(u, \theta)
  \sin \theta}{r \sin \theta}+O\left(\frac{1}{r^2}\right),
 \ea
 and the non-vanishing component of the basic vector is \be T^0 \cong 2\,\frac {1+M_{,\, 0}(
u,\theta)}{r}+O\left(\frac{1}{r^2}\right).\ee

\end{document}